\documentclass[floatfix,amssymb,prl,twocolumn,superscriptaddress,nofootinbib, aps]{revtex4-1}
		
\usepackage{amssymb,amsmath,verbatim,mathtools,needspace,enumitem,etoolbox,graphicx,physics,microtype,afterpage,bm,soul}
\usepackage[dvipsnames]{xcolor}
\definecolor{linkcolor}{rgb}{0.0,0.3,0.5}
\usepackage[unicode, colorlinks=true, linkcolor=linkcolor, citecolor=linkcolor, filecolor=linkcolor,urlcolor=linkcolor, pdfusetitle]{hyperref}
\usepackage{orcidlink}
\usepackage[all]{hypcap}
\usepackage[T1]{fontenc}
\usepackage[utf8]{inputenc}
\usepackage{tabularx}
\usepackage{float}
\usepackage{lmodern}
\usepackage{ragged2e}
\allowdisplaybreaks
\usepackage{tikz}
\usepackage{color}
\usepackage{framed}
\usepackage{hyperref}
\hypersetup{colorlinks, citecolor=darkblue, linkcolor=black, urlcolor=darkblue}
\definecolor{rossos}{cmyk}{0,1,1,0.55}
\definecolor{bluscuro}{rgb}{0.15, 0.2, .85}
\definecolor{bluchiaro}{cmyk}{1,.3,0.,0.1}
\definecolor{ForestGreen}{rgb}{0.13, 0.55, 0.13}
\definecolor{darkblue}{rgb}{0,0, 1.39}
\newcommand{\msun}{M_{\odot}}

\newcommand{\be}{\begin{equation}}
\newcommand{\ee}{\end{equation}}

\def\lsim{\mathrel{\rlap{\lower4pt\hbox{\hskip0.5pt$\sim$}}
    \raise1pt\hbox{$<$}}}         
\def\gsim{\mathrel{\rlap{\lower4pt\hbox{\hskip0.5pt$\sim$}}
    \raise1pt\hbox{$>$}}}         

\begin{document}

\title{Primordial black hole ringdown:  \\
The irreducible stochastic gravitational wave background   }

\author{Valerio De Luca\orcidlink{0000-0002-1444-5372}}
\email{vdeluca@sas.upenn.edu}
\affiliation{Center for Particle Cosmology, Department of Physics and Astronomy,
University of Pennsylvania, 209 South 33rd Street, Philadelphia, Pennsylvania 19104, USA}
\author{Antonio J. Iovino\orcidlink{0000-0002-8531-5962}}
\email{ai2869@nyu.edu}
\affiliation{New York University, Abu Dhabi, PO Box 129188 Saadiyat Island, Abu Dhabi, UAE}
\author{Antonio Riotto\orcidlink{0000-0001-6948-0856}}
\email{antonio.riotto@unige.ch}
\affiliation{D\'epartement de Physique Th\'eorique and Gravitational Wave Science Center (GWSC), Universit\'e de Gen\`eve, CH-1211 Geneva, Switzerland}


\begin{abstract}
\noindent
Independently from the formation mechanism of primordial black holes in the early Universe, their generation is accompanied by a ringdown phase during which they relax to a stationary configuration and gravitational waves under the form of quasinormal modes are emitted. Such gravitational waves generate an irreducible and unavoidable stochastic background which is testable by current and future experiments. In particular, for primordial black holes with masses exceeding $10^{14}\,M_{\odot}$, the associated stochastic background lies within the frequency range accessible to current and upcoming  cosmic microwave background experiments, thereby providing a direct observational way to probe the existence of such extremely massive objects.

\end{abstract}

\maketitle

\vspace{0.1cm}
\noindent{{\bf{\em Introduction.}}} In recent years, there has been a growing focus on the study of Primordial Black Holes (PBHs) (see~\cite{Sasaki:2018dmp, Carr:2020gox, Green:2020jor,LISACosmologyWorkingGroup:2023njw,Byrnes:2025tji,Carr:2026hot} for comprehensive reviews), spurred in large part by the various observations of gravitational waves (GWs) from black hole binary mergers~\cite{LIGOScientific:2016aoc, LIGOScientific:2018mvr, LIGOScientific:2020ibl, KAGRA:2021vkt}, and the proposition that some of these events might have a primordial origin~\cite{Bird:2016dcv, Sasaki:2016jop, Clesse:2016vqa,Mukherjee:2021ags,DeLuca:2020sae,DeLuca:2020agl,Franciolini:2021tla,Afroz:2024fzp}.

A number of theoretical frameworks and production mechanisms have been put forward as plausible explanations for the formation of PBHs. These scenarios, which include the gravitational collapse of cosmological perturbations from inflation, the collision of bubble walls from first-order phase transitions, and the collapse of domain walls and cosmic strings, predict diverse PBH mass spectra depending on their underlying assumptions (see~\cite{LISACosmologyWorkingGroup:2023njw} for a detailed discussion).

Independently from the formation mechanism of PBHs, their generation is unavoidably accompanied by a final  ringdown phase during which the 
spacetime oscillates before relaxing to  a stationary and static Schwarzschild or Kerr configuration.  Indeed, BH linear perturbation theory  predicts that the latest stage of the  BH formation, when the spacetime around it settles down to a  static Schwarzschild or Kerr metric, exhibits the emission of GWs with a  characteristic pattern of exponentially decaying oscillations,  known as quasinormal modes (QNMs) (for a review, see Ref.~\cite{Berti:2025hly}). Let us stress that this scenario differs from the quadrupolar emission of GWs associated to asphericities in the gravitational collapse~\cite{DeLuca:2019llr}.

In this Letter we calculate the stochastic GW background (SGWB)
generated during the ringdown phase of PBHs. Such a SGWB has to be regarded as irreducible, in the sense that it represents the unavoidable and minimal amount of GWs associated to PBH formation, whatever their masses and abundance are. Interestingly, we find that such a background could also be within the reach of current and upcoming cosmic microwave background (CMB) experiments, for extremely massive BHs in the mass range $10^{14}\, M_\odot$ to $10^{18}\, M_\odot$.\\
In the following we use geometrical  units and set $c = G_\text{\tiny N} = 1$.


\vspace{0.1cm}
\noindent{{\bf{\em GWs from PBH ringdown.}}} The ringdown phase produces GWs which, far enough from the BH location, have the following form for the (dimensionless) strain
\be
h_{\ell m}(u,r)=A_{\ell m}\frac{M}{ r}e^{- i \, \omega_{\ell m}u}\,,
\ee
when considering only the fundamental mode. Here, $u=(t-r)$ is the retarded time,  $(\ell,m)$ are the multiple numbers in a spherical harmonics decomposition, $A_{\ell m}$ is the dimensionless amplitude of the GW, and $\omega_{\ell m}$ is the corresponding QNM frequency.
The energy density associated to such an emission reads 
\begin{align}
E&=\frac{3}{4\pi}r^2\sum_{\ell m}\int_0^u\,{\rm d}u'\,\left|\frac{\partial h_{\ell m}}{\partial u'}\right|^2\nonumber\\
&=  \frac{3}{8\pi}M^2\sum_{\ell m}|A_{\ell m}|^2\frac{|\omega_{\ell m}|^2}{|{\rm Im}\,\omega_{\ell m}|}\,,
\end{align}
where we have accounted for the two GW degrees of freedom and integrated up  to times sufficiently longer than the Hubble time $H^{-1}$ at which the BHs are formed, $u\gg H^{-1}$. In the following we will make the conservative assumption of focusing only on the leading modes $\ell=m=2$, for which $\omega_{22}M\simeq (0.37-0.09 \, i)$, so that $E\simeq 0.2 |A|^2\, M$, in terms of the final BH mass $M$~\cite{Leaver:1986gd} (we have renamed $A_{22}=A$ to avoid cluttering notation). In the estimates above, we have used the QNM frequency of a Schwarzschild BH for simplicity. However, the analysis can be straightforwardly generalized to the case of a Kerr BH.

Inside an Hubble volume, the GW density is therefore
\be
\rho_\text{\tiny GW}\simeq \frac{3E}{4\pi H^{-3}} \beta \simeq 0.05\,|A|^2\, \beta\,M H^3\,,
\ee
where, assuming a PBH monochromatic distribution, we have included the factor~\cite{Carr:2009jm}
\be
\beta \simeq 6 \cdot 10^{-9}\left(\frac{\gamma}{0.2}\right)^{-\frac{1}{2}}\left(\frac{g_{*}}{106.75}\right)^{\frac{1}{4}}\left(\frac{M}{M_{\odot}}\right)^{\frac{1}{2}} f_\text{\tiny PBH}\,,
\ee
which indicates the probability to form one PBH within a Hubble volume in units of $f_\text{\tiny PBH}$, the current abundance  of PBHs as a fraction of the dark matter. In this expression $\gamma = 2MH$ (again in units of $G_\text{\tiny N}=1)$ indicates the fraction of PBH mass compared to the whole mass included in a Hubble volume~\cite{Carr:1975qj}, while $g_*$ denotes the number of relativistic degrees of freedom in the Standard Model. We set for simplicity $g_*=106.75$.

\begin{figure*}[t!]
    \centering
\includegraphics[width=0.99\textwidth]{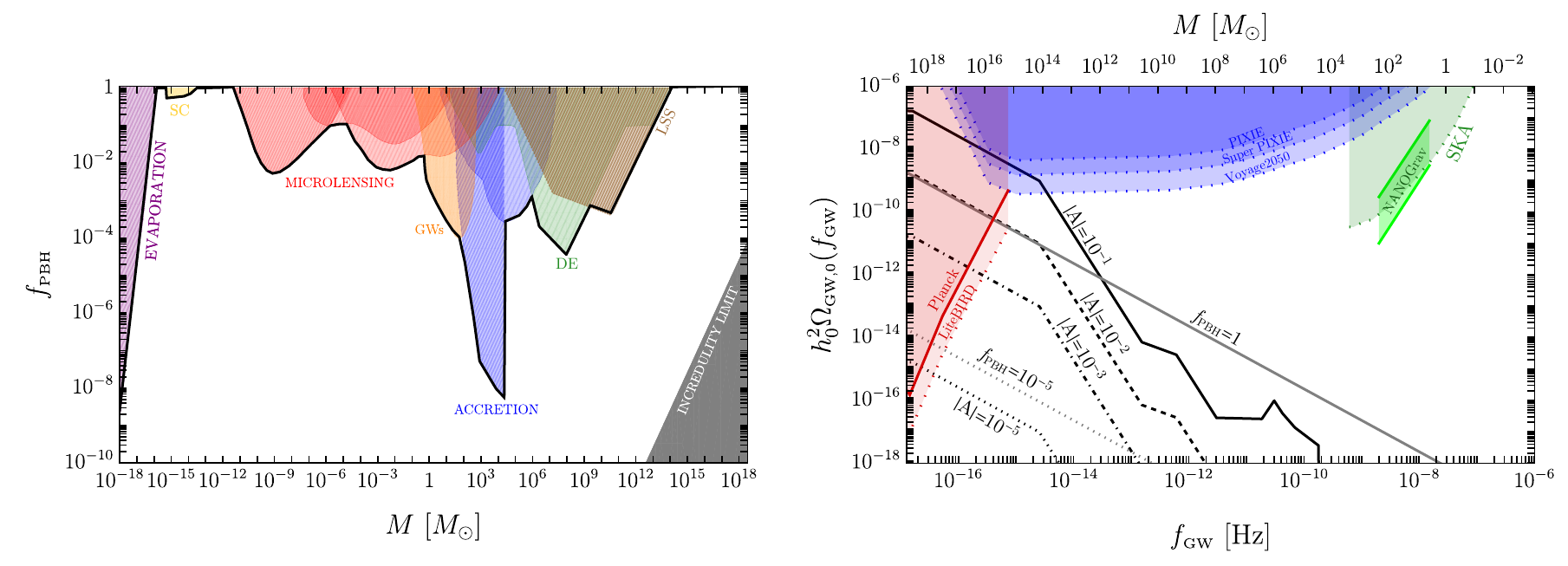}
    \caption{{\it Left panel}:  Current constraints on the PBH abundance for a monochromatic population of non-spinning PBHs (see the main text for a description). The maximum value of the PBH abundance, namely $f^{\text{\tiny max}}_\text{\tiny PBH}(M)$, is shown by a black line.
    {\it Right panel}: Prediction for the present GW abundance associated to the SGWB induced by PBH ringdown. The black lines denote the predictions assuming a QNM amplitude $|A|=( 10^{-5},10^{-3},10^{-2},10^{-1})$ and $ f_\text{\tiny PBH}= f^{\text{\tiny max}}_\text{\tiny PBH}(M)$, while solid (and dotted) gray lines are obtained fixing $|A|=10^{-2}$ and assuming $f_\text{\tiny PBH}=1$ $(10^{-5})$.  The plot also shows the constraints derived from ongoing experiments (solid colored lines) and sensitivity from future missions (dashed colored lines) (see the main text for a description). We stress that the black and gray lines should not be interpreted as a continuous signal but as a set of points; given a monochromatic population of PBHs with mass $M$, one can determine first the peak frequency at which the signal is produced, $f_\text{\tiny GW}$, and then the corresponding amplitude at the peak of the signal, $h_0^2\,\Omega_{\text{\tiny GW},0}(f_\text{\tiny GW})$.
    }
\label{fig:1}
\end{figure*}

The corresponding amount of GWs  in units of the critical energy density $\rho_c=3 H^2/8\pi$ at the time of emission then reads
\be
\Omega_\text{\tiny GW}=\frac{\rho_\text{\tiny GW}}{\rho_c}\simeq 0.4\,\beta\cdot\gamma\,|A|^2\,.
\ee
 One finds that the amount of GWs today is
\be\label{eq:omtot}
 h_0^2\,\Omega_{\text{\tiny GW},0} \,= c_g h_0^2 \, \Omega_r \, \Omega_\text{\tiny GW} \simeq 6.7\cdot 10^{-6}\,\beta\cdot\gamma\,|A|^2\,,
\ee
where $h_0\simeq 0.75$ is the current Hubble constant in units of 100 km/sec/Mpc, $h_0^2 \Omega_r \simeq 4.2 \cdot 10^{-5}$ is the radiation density and $c_g \simeq 0.4$ takes into account the change in the number of degrees of freedom~\cite{Bartolo:2018evs, Bartolo:2018rku}.
Working under the assumption of a monochromatic population of PBHs, we can safely assume that our GW signal is nearly monochromatic, i.e. the value of the peak of GWs today, $h_0^2\,\Omega_\text{\tiny GW,0}(f_\text{\tiny GW})$, is well approximated by Eq.~\eqref{eq:omtot}. 

The frequency of the peak of the GWs associated to such background is dictated by the fundamental QNM frequency,  $|\omega_{22}|/2\pi \simeq 12\cdot(\msun/M)$ kHz, which, when redshifted today, reads
\be
\label{fGWM}
f_\text{\tiny GW}\simeq 2.9\cdot10^{-8} \left(\frac{\gamma}{0.2}\right)^{-1/2}
\left(\frac{M_\odot}{M}\right)^{1/2}\,{\rm Hz}\,.
\ee
The abundance of GWs at the peak frequency today can then be recast as 
\be
\label{SGWB}
h_0^2\,\Omega_{\text{\tiny GW},0} (f_{\text{\tiny GW}}) \, \simeq  2\cdot10^{-17}\,\left(\frac{|A|}{10^{-2}}\right)^2\,f_\text{\tiny PBH}\,\left(\frac{10^{-9}\,{\rm Hz}}{f_\text{\tiny GW}}\right)\,.
\ee
This equation provides the final estimate for the SGWB associated to the PBH ringing.

There is a final missing ingredient in the expression, given by  the typical value of the QNM amplitude $A$.
Independently from the PBH formation mechanism,  before the static Schwarzschild or Kerr configuration is reached, the spacetime metric is still time dependent, and the relaxation to the  static configuration results in the production of QNMs~\cite{Sun:1990pi}. In principle, one might study the properties of the QNMs and their amplitudes using a Vaidya-like metric~\cite{Vaidya:1951fdr, Nolan:2005is, Nolan:2006pz,Capuano:2024qhv}, describing a (non-empty) spherically symmetric BH spacetime with a dynamical mass $M(t)$, which may vary between zero and the final value $M$ in a given time range. In such a case, one expects $A$ to be proportional to $\dot M(t)$, thus not being far from unity, since a BH of mass comparable to $M \sim 1/H$ is formed within a Hubble time $1/H$. In our estimates we will thus use $|A|\simeq \mathcal{O}(10^{-3} \divisionsymbol 10^{-1})$ as a reasonable figure to estimate the SGWB and compare it with the sensitivity of various GW experiments, as we will show in the following.\footnote{As an explicit example, in the case of PBH formation from the collapse of Gaussian inflationary density peaks, the GW amplitude parameter scales as
$A \sim Q H^2/M \sim e$, where $Q$ is the quadrupole moment of the collapsing ellipsoidal region of size $\sim 1/H$ and $e$ denotes the ellipticity of the peak. For Gaussian primordial fluctuations, the statistics of peaks \cite{BBKS} give a mean ellipticity
$\langle e \rangle = 1/(\sqrt{5} \gamma\nu)$, where $\gamma \sim \mathcal{O}(1)$ and $\nu$ is the peak height. For rare peaks relevant to PBH formation ($\nu \sim 10$), this yields a typical estimate $A \sim \mathcal{O}(10^{-1})$.} 
Let us also stress that the QNM frequencies would be slightly modified when considering a Vaidya rather than a static background metric~\cite{Capuano:2024qhv}. However, we expect that our conclusions would not be significantly  affected as well as if we consider a Kerr BH \cite{Berti:2005ys}.

\vspace{0.1cm}

\noindent{{\bf{\em Detectability.}}}
To predict the SGWB associated with PBH ringdown, as shown in Eq.~\eqref{SGWB}, it is necessary to account for current constraints on the PBH dark matter fraction, $f_\text{\tiny PBH}$, derived from a wide range of astrophysical and cosmological observations. These constraints, shown in the left panel of Fig.~\ref{fig:1} for a non-spinning monochromatic PBH mass distribution, can be categorized as follows: at low masses ($\lesssim 10^{-17}\, M_\odot$), PBH evaporation via Hawking radiation is constrained by the gamma-ray background, Big Bang nucleosynthesis, CMB spectral distortions and Lyman-$\alpha$ data (magenta) (see Refs.~\cite{Saha:2021pqf,Laha:2019ssq,Ray:2021mxu,Mittal:2021egv,Clark:2016nst,Laha:2020ivk,Berteaud:2022tws,DeRocco:2019fjq,Dasgupta:2019cae,Calza:2021czr,Boudaud:2018hqb,Khan:2025kag}). At slightly heavier masses, a small parameter-space island is constrained by the capture of PBHs by stars in ultra-faint dwarf galaxies (constraints labeled SC, yellow)\,\cite{Esser:2025pnt}. In the intermediate range ($\sim 10^{-11}\, M_\odot$ to a few tens of solar masses), microlensing surveys limit the PBH abundance from the non-detection of lensing events (red) (see Refs.\,\cite{Niikura:2017zjd,Niikura:2019kqi,Mroz:2024wag,Mroz:2024wia}). Around stellar masses, LIGO-Virgo-KAGRA GW detections constrain PBHs by comparing the observed merger rate with PBH merger predictions (orange) (see  Refs.\,\cite{LIGOScientific:2019kan,Kavanagh:2018ggo,Wong:2020yig,Hutsi:2020sol,DeLuca:2021wjr,Franciolini:2022tfm,Andres-Carcasona:2024wqk}). 
For PBHs above $\sim 1\,M_\odot$, constraints are obtained due to the modification of the CMB through enhanced ionization and spectral distortions from baryonic accretion and X-ray binaries (blue) (see Refs.\,\cite{Serpico:2020ehh,Agius:2024ecw,Manshanden:2018tze}). At higher masses ($\gtrsim 10^2\,M_\odot$), dynamical constraints arise from the disruption of wide binaries, the stability of globular clusters, and the survival of dwarf galaxies (green) (see Refs.\,\cite{Carr:2018rid,Carr:2020erq}).Finally, for very massive PBHs ($\gtrsim 10^7\,M_\odot$), limits come from their impact on large-scale structure formation and galaxy clustering (brown) (see Refs.~\cite{Carr:2018rid,Carr:2020erq}).
The gray region in the bottom right corner shows the incredulity limit\,\cite{Carr:2020erq}, which corresponds to having fewer than one single PBH within the current Hubble volume.

Taking into account the upper bound on $f_{\text{\tiny PBH}}$ provided by the above constraints for each PBH mass, namely $f^{\text{\tiny max}}_\text{\tiny PBH}(M)$ (shown as a black line in the left panel of Fig.~\ref{fig:1}), one can estimate Eq.~\eqref{SGWB} in terms of the corresponding GW frequency by using the relation given in Eq.~\eqref{fGWM}.
The result is shown as the black and gray lines in the right panel of Fig.~\ref{fig:1}. These identify the irreducible SGWB associated to the ringing of newly formed PBHs, considering  optimistic or more realistic QNM amplitudes from $|A| \simeq \mathcal{O}(10^{-1})$ to $|A| \simeq \mathcal{O}(10^{-5})$, and compatible with current constraints on their abundance.

These predictions can be compared with the sensitivity curves of present (solid lines) or future (dashed lines) experiments looking for GW signals. These include current constraints from NANOGrav 15 yrs experiment\,\cite{NANOGrav:2023gor,NANOGrav:2023hvm} and sensitivity from future pulsar timing array (PTA) experiments such as Square-Kilometer Array (SKA)\,\cite{Zhao:2013bba,Babak:2024yhu} (green); sensitivity from future proposed CMB experiments looking for B-modes, such as PIXIE\,\cite{Kogut:2011xw}, Super Pixie\,\cite{Kogut:2019vqh} and Voyage2050\,\cite{Chluba:2019nxa} (blue); and current excluded region (Planck\,\cite{Planck:2018jri,BICEP2:2018kqh,Tristram:2020wbi,BICEP:2021xfz}) and future sensitivity (LiteBIRD\,\cite{Hazumi:2019lys}) from space-based experiments targeting the indirect detection of primordial gravitational waves through the measurement of the CMB B-modes.

The figure illustrates that, although the predicted signal remains several orders of magnitude below the current sensitivity limits of PTA experiments, it can still lie within the constraints already set by Planck. Moreover, upcoming CMB missions such as LiteBIRD, Super PIXIE and Voyage2050 will have the capability to probe this unavoidable contribution to the SGWB with unprecedented precision, and, more generally, to place meaningful constraints on the abundance of PBHs in the very high mass range. Indeed, a non-detection of such a signal places meaningful upper limits on the abundance of ``stupendously large'' BHs (SLABs) in the Universe, further tightening the constraints on this elusive population.\footnote{Recently, several works have appeared aiming to set additional constraints in the mass range of interest~\cite{Ivanov:2025pbu,Gerlach:2025vco}.}
Clearly, the exact lower bound on $f_{\text{\tiny PBH}}$ imposed by these constraints intrinsically depends on the value of $|A|$ and, for this reason, it cannot be precisely determined within the SLAB mass range. However, Planck constraints already rule out the possibility that SLABs constitute the entirety of dark matter across the entire mass range --- that is, $f_{\text{\tiny PBH}} = 1$ for $M \gtrsim 10^{15}\, M_\odot$ --- if $|A| \gtrsim 10^{-1.5}$. In contrast, the window in this extreme mass range remains completely open for $|A| \lesssim 10^{-5.5}$.

\vspace{0.1cm}

\noindent{{\bf{\em Discussion and conclusions.}}}
In this Letter we have pointed out  that any mechanisms for the production of PBHs in the universe, comprising a sizable fraction of the dark matter, are responsible for the generation of an irreducible SGWB associated to the QNM ringing of these objects when they relax to a stationary configuration after formation. 

Given its irreducible nature, this signal can be used to set further robust constraints on the abundance or, in the future to determine the existence, of extremely massive BHs, often referred to as  SLABs~\cite{Carr:2020erq}, typically defined in the mass range above $10^{12}\,M_\odot$. These exotic objects could originate from a variety of formation mechanisms (see, {\it e.g.}, Ref.~\cite{Faria:2025nlm} for an overview of proposed scenarios) and are subject to constraints from several astrophysical and cosmological observables~\cite{Carr:2020erq, Deng:2021edw,Gerlach:2025vco}.  The formation and subsequent evolution of these objects would unavoidably generate a SGWB associated with their ringdown phase. Current and future CMB experiments, with their improved sensitivity to low-frequency GWs, will thus offer a unique opportunity to probe this background. A non-detection of such a signal places meaningful upper limits on the abundance of SLABs in the universe, further tightening the constraints on this elusive population. Clearly, the exact lower bound on $f_{\text{\tiny PBH}}$ imposed by these constraints intrinsically depends on the value of $|A|$ and, for this reason, it cannot be precisely determined within the SLAB mass range.

The unavoidable nature of the ringdown signal, present for any PBH formation scenario, implies that it will add to the GW backgrounds already expected for each specific formation mechanism. For example,  if additional lighter PBHs are formed from the gravitational collapse of large curvature perturbations generated during inflation, one expects an additional second-order scalar-induced GW background (SIGW)~\cite{Matarrese:1993zf,Acquaviva:2002ud, Mollerach:2003nq,Ananda:2006af,Baumann:2007zm,Domenech:2021ztg}, with an amplitude (evaluated at the peak frequency, for a monochromatic primordial power spectrum) given by $\Omega_\text{\tiny SIGW} \, h_0^2 \simeq 10^{-5} A_\text{\tiny PS}^2$~\cite{Kohri:2018awv}, 
where $A_\text{\tiny PS}$ denotes the amplitude of the primordial power spectrum at the relevant scale.
While at frequencies in the nHz range this SIGW contribution can lie within the sensitivity bands of current and future GW experiments (see, e.g., Refs.~\cite{Chen:2019xse,DeLuca:2020agl,Vaskonen:2020lbd,Kohri:2020qqd,Domenech:2020ers,Dandoy:2023jot,Franciolini:2023pbf,Wang:2023ost,Liu:2023ymk,Iovino:2024tyg,Esmyol:2025ket, Cyr:2023pgw}), at lower frequencies corresponding to more massive PBHs---including SLABs---the amplitude $A_\text{\tiny PS}$ is strongly constrained by observations of CMB $\mu$-distortions and Lyman-$\alpha$ forest data to be $A_\text{\tiny PS} \lesssim 10^{-9}$~\cite{Planck:2018jri,Bird:2010mp}. This implies a SIGW energy density as small as $\Omega_\text{\tiny SIGW} \, h_0^2 \approx 10^{-23}$, which is several orders of magnitude below the expected ringdown contribution. Similarly, if PBHs are produced through the collisions of bubble walls during a first-order phase transition, the ringdown contribution will add to the stochastic GW spectrum associated with the phase transition itself~\cite{Lewicki:2023ioy, Lewicki:2023mik, Lewicki:2024ghw, Lewicki:2024sfw,Franciolini:2025ztf}. A detailed quantitative comparison between the SGWB from PBH ringdown and the GW signals predicted for each specific PBH formation scenario or tensor modes generated during inflation is left for future work.

There are several other promising directions for future research. An immediate priority is to determine the amplitude of the QNM spectrum through dedicated numerical relativity simulations of gravitational collapse, enabling a more accurate prediction of the resulting SGWB. Additionally, it would be valuable to incorporate further physical effects in the modeling of the emitted GW signal, such as tail effects
arising from back-scattering on the PBH spacetime or nonlinear contributions from QNM interactions and GW memory \cite{Price:1972pw}.
Furthermore, our computation can be extended to the case of a broad PBH mass function and to scenarios with different values of the cosmic equation-of-state parameter, which would result in a broader GW ringdown spectrum across frequencies~\cite{Carr:2019kxo,Carr:2026hot}.
Finally, given that PBHs are expected to form within external environments, accounting for environmental effects on the QNM spectrum represents another important extension~\cite{Barausse:2014tra,Cardoso:2021wlq,Destounis:2022obl,Cardoso:2022whc,Ianniccari:2024ysv,Spieksma:2024voy}.
We leave the exploration of these aspects to future work.

\vspace{0.1cm}
\noindent{{\bf{\em Acknowledgments.}}}
We thank B. Carr, G. Franciolini, G. Perna, D. Perrone, F. Quevedo and K. Schmitz for useful discussions and comments on the draft.
V.DL. is supported by funds provided by the Center for Particle Cosmology at the University of Pennsylvania. 
A.J.I. acknowledges the University of Trento for the nice hospitality during the realization of this project.
A.R. acknowledges support from the Swiss National Science Foundation (project number CRSII5 213497) and from the Boninchi Foundation for the project “PBHs in the Era of GW Astronomy”.
\bibliography{draft}

\end{document}